\documentstyle[epsfig]{article}
\newcommand{\lsim}{\:\raise -4pt\hbox{$\stackrel{\textstyle <} {\sim}$}\:}
\begin{document}

\begin{center}
\large
{\bf LUCIAE 3.0: A new version of a computer program for Firecracker Model
 and rescattering in relativistic heavy-ion collisions}\\
\vspace{0.3in}
\vspace{1cm}\large
  Tai An$^{1,2}$ and Sa Ben-Hao$^{1,3,4}$ \\
\begin{tabbing}
ttttt \= tt \= \kill
\>1. CCAST (World Lab.), P. O. Box 8730 Beijing, 100080 \\
\> \>China. \\
\>2. Institute of High Energy Physics, Academia Sinica, \\
\> \> P. O. Box 918, Beijing, 100039 China.\footnotemark \\       
\>3. China Institute of Atomic Energy, P. O. Box 275, \\
\> \>Beijing, 102413 China.\\
\>4. Institute of Theoretical Physics, Academia Sinica, \\
\> \>Beijing 100080 China.\\

\end{tabbing}
\end{center}
\footnotetext{mailing address. \\Email: taian@hptc1.ihep.ac.cn}

\normalsize
\large

Abstract: 
LUCIAE is a Monte Carlo program that, connected to FRITIOF, implements
both the Firecracker Model (FCM), a possible mechanism for collective
multi-gluon emission from the colour fields of interacting strings, 
and the reinteraction of the final state hadrons in relativistic 
heavy ion collisions. This paper includes a brief presentation of the
dynamics of LUCIAE with an emphasis on the new features in this version,
as well as a description of the program.

PACS number: 25.75.-q, 24.10.Lx, 25.75.Dw
\newpage
\section*{NEW VERSION SUMMARY}

\normalsize

{\it Title of Program:} LUCIAE version 3 \\

{\it Program obtainable from:} taian@hptc1.ihep.ac.cn or sabh@mipsa.ciae.ac.cn
 or from the CPC Program Library, Queen's University of Belfast, N. Ireland\\
 
{\it Reference in CPC to previous version:} 
Sa Ben-Hao and Tai An, Computer Physics Commun. 
90 (1995) 121 (version 2.0)

{\it Does the new version supersede the previous one:} yes

{\it Computer for which the programme is designed:} HP, ALPHA and DEC station,
VAX, IBM and others with a FORTRAN 77 compiler \\

{\it Computer:} HP station Model 715/100; installation: Computer Center, Institute of  
High Energy Physics, Beijing, China\\

{\it Operating system:} HP$-$UX 10.20

{\it Program language used:} FORTRAN 77 \\

{\it High speed storage required:} $\approx 90$k words\\

{\it No. of bits in a word:} 32 \\

{\it Peripherals used:} terminal for input, terminal or printer
for output \\

{\it No. of lines in combined program and test deck:} about 12100 \\

{\it Keywords:} Relativistic nucleus-nucleus collisions, Monte Carlo, 
collective effects, Colour Rope, effective string tension, rescattering. \\

{\it Nature of the physical problem:} The experiments of relativistic pA and 
AA collisions reveal that high energy heavy-ion collisions have some
features which can not be understood by the simple 
superposition of independent nucleon-nucleon collisions. They indicate clearly 
that the collective effects are important in the 
relativistic pA and AA collisions. Formation of a QGP is often 
suggested to be a candidate to account for some of these collective effects.
Could we understand those new features in pA and AA collisions by 
conventional physics (on which LUCIAE is based) ? What is the limit of a 
model to explain the experimental data without the formation of QGP  
within a reasonable margin of flexibility of the model ? The Monte-Carlo
generator, LUCIAE is built in an attempt to answer these questions.\\

{\it Method of solution:} 
When many strings or colour dipoles are formed in relativistic pA and 
AA collisions, 
it is natural to ask if there is some 
interaction among those strings close by so that both the
emission of gluonic bremsstrahlung as well as the fragmentation properties 
can be affected by the large common energy density of the string cluster 
(Colour Rope). The Firecracker Model is developed to study such a collective 
effect. Moreover, many hadrons are produced in a small space-time volume 
through fragmentation of these strings, which implies that they will interact 
with each other and with the surrounding cold spectator matter. The 
rescattering effect on the distributions of the final state hadrons
is also included in LUCIAE program.

{\it Summary of revisions:}

(1) The initialization of hadrons in the space-time is modified.

(2) Much more inelastic channels and their reverse reactions  have been 
included in the rescattering sector. 

(3) The annihilation of $\bar{p}$
and $\bar{\Lambda}$ in nuclear matter is taken into account. 

(4) The effect of firecracker gluons on the fragmentation of a string has been 
included.
 
(5) The sizes of arrays have been enlarged; The values of a few
parameters are adjusted and a few bugs are fixed.

{\it Restriction of complexity of the problem:}
 At very high energies ($\sqrt{s}$ in the TeV range), especially
 for collisions of massive nuclei, certain arrays need to be expanded to
 accommodate the large number of particles produced.   

{\it Typical running time:} Depends on the type of collision and energy.  
Three examples of central collisions (b=0) running on a HP Station:
\begin{tabbing}
 $^{16}$O $+$ $^{197}$ Au \qquad \= $p_{\rm lab} = 200$ A GeV/$c$:\qquad \=
 2 events/min (central collision)      \kill
$^{32}$S $+$ $^{32}$S 
\> $p_{\rm lab} = 200$ A GeV/$c$: \> 
$\sim$ 0.6 minutes/event \\
$^{207}$Pb $+$ $^{207}$Pb \> $p_{\rm lab} = 158$ A GeV/$c$: \> 
$\sim$ 9 minutes/event \\
$^{197}$Au $+$ $^{197}$Au \> $\sqrt{s} = 200$ GeV: \> 
$\sim$ 80 minutes/event    
\end{tabbing}
\vspace{0.3in}


\newpage
\large
\section*{LONG WRITE-UP}

\section{Introduction} 
With the advent of the BNL RHIC experiment the center of relativistic
heavy-ion collisions is moving to the study of RHIC physics for Au + Au
collisions at $\sqrt s$=200 GeV. At such a high energy 
heavy-ion collisions are expected to demonstrate many new features which
are not covered at current AGS and CERN energies.
First of all, the energy density created after the collisions
is expected higher than its critical value ($1-3\  GeV/fm^3$ 
by lattice calculation \cite{lattice}) of a phase transition from
a hadronic state to a QGP state. Second, QCD hard and semi-hard 
processes will become important and affect the evolution of the 
colliding system \cite{glu}. \\

The study of relativistic heavy-ion collisions at AGS and CERN energies, 
though a QGP phase transition is not unambiguously seen, demonstrates that
pA and AA collisions are far more than independent superposition of
binary nucleon-nucleon collisions. Many interesting collective effects
occur during the collisions, some of which have thus been considered
as signatures of the formation of a QGP state. In order to distinguish
the real QGP signatures from ``conventional'' collective effects it is 
important to develop models based on the different non-QGP collective effects.
 Such models should be used in order to provide possible 
background for the signals of QGP formation.

FRITIOF is a string model specially developed for simulating relativistic 
hadron-hadron scattering by the introduction of gluon
bremsstrahlung radiation as well as hard parton scattering and it has
been successful in describing many experimental data at the SPS energies, 
which shows that the production of QCD hard and semi-hard jets plays a 
crucial role at such high energies\cite{fritiof}. On the basis 
of hadron-hadron interaction the FRITIOF model has also been extended 
to describe hadron-nucleus and nucleus-nucleus collisions 
in a way similar to Glauber model. 

In the FRITIOF string language two strings are formed during a
high energy hadron-hadron collisions. They pass by each other quickly
and have little chance for further interaction. However, there are 
generally many
excited strings formed close by each other during a relativistic
heavy-ion collision. In case these strings would behave like vortex
lines in a color superconducting QCD vacuum then it is conceivable that
they will interact. The interacting string may form 
a quantum state (``Color Rope'') so that both the
emission of gluonic bremsstrahlung as well as the fragmentation
properties can be affected by the large common energy density.
Such a scenario is described by ``Firecacker model'' and its application to 
the production of mini-jets has been discussed in references \cite{fire}.

Besides, many hadrons are produced 
in a small space-time volume through the string fragmentation  
in a relativistic heavy-ion collision, which implies that they 
would interact with each other and with the surrounding 
cold spectator matter. Such rescattering turns out to be important in the
production of strange particles and anti-baryons during a collision
\cite{rescat1}.

The above collective effects, the production of firecracker gluons and
rescattering of the final state hadrons, has been implemented in a 
computer program, LUCIAE\_2.0 \cite{luciae}, and we have 
kept on developing the program ever since. The following changes
have been made in the new version of the program, LUCIAE\_3.0: (1) the 
initialization of hadrons in the space-time is modified so that 
uncertainty due to the initialization is minimized.
(2) much more inelastic channels and their reverse reactions  have been 
included in the rescattering sector so that the program can be used to 
investigate production of those strange particles like $\Xi, \Omega^-$, for 
instance, through rescattering. (3) The annihilation of $\bar{p}$
and $\bar{\Lambda}$ in nuclear matter is taken into account. 
(4) the effect of firecracker gluons on the fragmentation of a string has been 
included through the change of effective string tension in the new
version of the program. (5) The sizes of arrays have been enlarged
for simulating Au+Au at the RHIC energy; The values of a few
parameters are adjusted and a few bugs are fixed.
The physics implemented in LUCIAE\_3.0 has been published
in papers \cite{rescat2}\cite{rescat3}.

In this paper we will briefly summarize the physics on which LUCIAE program
is builded with an emphasis on the new features in LUCIAE\_3.0. 
Firecracker Model is described in Section \ref{FCM}. The rescattering part
is described in Section \ref{Rescatt}. The description of the MC
program as well as its switches and parameters will be found in Section 
\ref{descript}.
\section{Firecracker Model}
\label{FCM}
When a gluon is emitted from a color dipole string with a mass $M$ and 
transverse size 
of an order of $1/\mu$ in the cms of the string the transverse 
momentum  $k_{\perp}$ and rapidity $y$ of the emitted gluon are 
restricted inside the region described in the soft radiation 
model\cite{SRM} by
\begin{equation}
k_{\perp} \lsim e^{-|y|} \frac{\mu}{k_{\perp}} M.
\label{f11}
\end{equation}
We note
that the factor $\mu M$ in eq. (\ref{f11}) effectively corresponds to
an energy density over the region $1/\mu$. Consequently the 
requirement in eq. (\ref{f11}) means for an isolated string that the
largest transverse momentum gluon fulfills $k_{\perp max} \leq \sqrt{\mu M}$.

In the Firecracker model it is assumed that in the very early stage of the 
collision when several ($n$) excited strings (energy momenta $P_{\pm j}$) are
close by {\em it may be possible for them to emit gluons using their
common total energy density} $\mu M_{tot} \equiv \mu \sqrt{\sum
P_{+j}\sum P_{-j}}$.

Consequently, when it comes to heavy ion collision predictions the Firecracker
model will correspond to an essential enhancement of (mini)jets in the center 
of phase space, which contributes to high $p_{t}$ enhancement. Actually the 
Firecracker gluons are found so dominant in the high $p_{t}$ region of phase 
space that the effect completely drowns both the Rutherford scattering and the 
``ordinary'' bremsstrahlung at RHIC energy. For more details how to combine 
strings close by each other transversely into a colur rope and how to 
partition a firecraker gluon to a string, see reference \cite{luciae}.

The production of the firecracker gluons corresponds to a sizable transverse 
excitement ( a `` kink " ) on a string, one of effects of which is that the
existence of the firecracker gluons on a string will
wrinkle the string and give a fractal structure. Such a wrinkled string has 
larger energy density in comparison with a string without gluon, thereby an 
enhanced string tension effectively \cite{torbjon}.\\

The following form has been used in \cite{rescat3} to parametrize the relation
between the effective string tension and the hard gluon jets on a string 
 \begin{equation}
\kappa_{eff}=\kappa_{0} (1-\xi)^{-\alpha},
\label{f1}
\end{equation} 
where $\kappa_{0}$ is the string tension of the pure $q\bar{q}$ string, 
$\alpha$ is a parameter to be determined by experiments and $\xi$ is 
calculated by
\begin{equation}
\xi =\frac{\ln(\frac{k_{\perp max}^2}{s_{0}})}{\ln (\frac{s}{s_{0}}) + 
\sum_{j=2}^{n-1} \ln (\frac{k_{\perp j}^2}{s_{0}})},
\label{f2}
\end{equation}
which represents the scale that a multigluon string is deviated from a pure 
$q\bar{q}$ string. Here the multigluon string state has (n-2) gluons, indexed 
in a colour connected way from the $q$ (index 1) to the $\bar{q}$ (index n) 
and $k_{\perp j}$ are the transverse momenta of the emitted gluons with 
$k_{\perp j}^2 \geq s_{0}$. The parameter $\sqrt{s_{0}}$ is of the order of a 
typical hadron mass. The parameter $\alpha$ in Eq.(\ref{f1}) and the 
$\sqrt{s_{0}}$ in Eq.(\ref{f2}) are determined by hh data to be about 3.5 and 
0.8 GeV, respectively \cite{rescat3}.

In the Lund string fragmentation model, the $q\bar{q}$ pairs with the quark 
mass $m$ and the transverse momentum $p_{t}$ are produced from the colour 
field by a quantum tunneling process with probability
\begin{equation}
\exp(\frac{-\pi m^{2}}{\kappa_{eff}})\exp(\frac{-\pi p_{t}^{2}}{\kappa_{eff}}).
\label{f5}
\end{equation}
The above equation shows that the probability of the $s\bar{s}$ pair 
production with respect to a $u\bar{u}$ (or $d\bar{d}$) pair as well as the 
probability of a high $p_{t}$ $q\bar{q}$ pair production will be enhanced in a 
field with larger $\kappa_{eff}$. 

Assume that the width of the Gaussian transverse momentum distribution 
and the strangeness suppression factor of a string with effective string 
tension $\kappa_{eff1}$ are $\sigma_{1}$ and $\lambda_{1}$, respectively, then
those quantities of a string with effective string tension $\kappa_{eff2}$ can 
be calculated from Eq.(\ref{f5}), i.e.  
\begin{eqnarray}
\sigma_{2} &=& \sigma_{1}(\frac{\kappa_{eff2}}{\kappa_{eff1}})^{1/2}\nonumber\\
\lambda_{2} &=& \lambda_{1}^ {\frac{\kappa_{eff1}}{\kappa_{eff2}}}.
\label{f6}
\end{eqnarray}

We see that $\sigma$ and $\lambda$ for two string states are related by the 
ratio of the effective string tensions of this two string states only. 
It should be noted that the discussion above is also valid for the production 
of the diquark pairs from the string field, i.e. the production of the diquark 
pairs with respect to the $q\bar{q}$ pairs will be enhanced from a string
with larger $\kappa_{eff}$, therefore, more baryons (or antibaryons) will be 
formed in the final state.

In JETSET routine which runs together with LUCIAE event generator, there are 
model parameters PARJ(2) (the same as $\lambda$) and PARJ(3), both of which 
are responsible for the s quark (diquark) suppression and related to the 
effective string tension. PARJ(3) is the extra suppression of strange diquark 
production compared to the normal suppression of strange quark pair. Besides 
$\lambda$ and PARJ(3) there is PARJ(1), which stands for the suppression of 
diquark - antidiquark pair production in the color field in comparison with 
the quark - antiquark pair production and is related to the effective string 
tension as well. How these three parameters affect the multiplicity 
distribution of final state particles can be found in \cite{rescat2}. Another 
parameter PARJ(21) (the same as $\sigma$), which is the width of the Gaussian 
transverse momentum distribution of $q\bar{q}$ pairs in the string 
fragmentation, varies with $\kappa_{eff}$ too, but it is not related to the 
strangeness production directly.

It has been shown in \cite{rescat2} that the string fragmentation by JETSET 
with default values of PARJ(1)=0.1, PARJ(2)=0.3 and PARJ(3)=0.4, determined 
from $e^+e^-$ experiments, overestimates the yield of strange particles in the 
pp collision at 200 GeV/c. Thus in \cite{rescat3} we first retune these 
parameters by comparing with the pp data of strange particle production. A 
new set of parameters PARJ(1)=0.046, PARJ(2)=0.2 and PARJ(3)=0.3 are found for 
pp at 200 GeV/c. We also give a new value of 0.32 for PARJ(21) (the 
corresponding default value is 0.37). This set of parameters are used to
calculate the particle production in pA and AA collisions at 200 GeV/c.

\section{Recattering of final state hadrons}
\label{Rescatt}
There is no space-time coordinate for the hadrons produced from strings
in FRITIOF. We therefore have to initialize these particles in space-time
in order to proceed the rescattering process. In LUCIAE the time origin of 
rescattering is chosen to be the moment when the distance 
between the center of the target and projectile along the beam 
direction is zero. It is then assumed 
that the particles from the string fragmentation are initially randomly
placed in space inside the geometrical overlap
region of the projectile and target nuclei. The ``spectator'' nucleons
are likewise randomly distributed outside the overlap region and inside the 
projectile (target) nucleus and provided with a thermal motion in accordance 
with a Boltzman distribution. We require as always that there is a minimum
distance between such spectators (``the hard intranuclear core
potential'') of $0.5$ fm. A formation time is given to each particle
and a particle starts to scatter with others after it is
``born''. The rescattering is performed in LUCIAE\_3.0 in the cms frame 
of two colliding nuclei (however, output of rescattering is still in the Lab 
frame). Due to the Lorentz contraction a relativistic heavy-ion collision in 
the simulation looks like what is described in 
the Bjorken model in which the collision happens between 
two thin disks and a particle will be ``born'' at a space-time point 
$z=\tau\sinh y , t=\tau\cosh y$ with $\tau$ and $y$ being the formation 
time and rapidity of the particle, respectively 
and z-axis along the beam direction. 
 
Two particles will collide if their minimum distance $d_{min}$ satisfies
\begin{equation}
\label{dmin}
d_{min} \leq \sqrt{\frac{\sigma_{tot}}{\pi}},
\end{equation}
where $\sigma_{tot}$ is the total cross section in fm$^2$ and the 
minimum distance is calculated in the cms frame of the two
colliding particles. If these two particles are moving towards each other 
at the time when both of them are ``born'' the minimum distance 
is defined as the distance perpendicular to the momentum of both
particles. If the two particles are moving back-to-back the 
minimum distance is defined as the distance at the moment when 
both of them are ``born''. Assuming that the hadrons move along 
straight - line classical trajectories between two consecutive collisions
it is possible to calculate the collision time when two hadrons 
reach their minimum distance and order all the possible collision 
pairs according to the collision time sequence.

If the total and the elastic cross section satisfies 
\begin{equation}
\frac{\sigma_{el}}{\sigma_{tot}} \geq \xi
\end{equation}
where $\xi$ is a random number, then the particles will be elastically
scattered or else the collision will be considered as an inelastic reaction.
The distribution of the momentum transfer, $t$, is taken as, 
\begin{eqnarray}
\frac{d\sigma}{dt} &\sim &\exp(Bt) ,
\end{eqnarray} 
where $B$, for an elastic scattering, depends on the masses of two
scattering particles. The azimuthal angle will be isotropically 
distributed.

The following inelastic reactions are included in LUCIAE3.0:
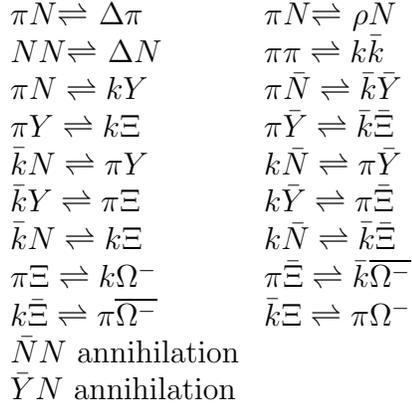
\begin{tabbing}
ttttttttttttttt\=ttttttttttttttt\=tttttt\=tttttttttttttttt\=  \kill
\>$\pi$$N$$\rightleftharpoons$ $\Delta$$\pi$
\> \>$\pi$$N$$\rightleftharpoons$ $\rho$$N$\\
\> $N$$N$$\rightleftharpoons$ $\Delta$$N$
\> \>$\pi\pi \rightleftharpoons k\bar{k}$\\
\>$\pi N \rightleftharpoons kY$
\> \>$\pi\bar{N} \rightleftharpoons  \bar{k}\bar{Y}$\\
\>$\pi Y  \rightleftharpoons k\Xi$
\> \>$\pi\bar{Y}  \rightleftharpoons  \bar{k}\bar{\Xi}$\\
\>$\bar{k}N  \rightleftharpoons  \pi Y$ 
\> \>$k\bar{N}  \rightleftharpoons  \pi\bar{Y}$\\
\>$\bar{k}Y  \rightleftharpoons  \pi\Xi$
\> \>$k\bar{Y}  \rightleftharpoons  \pi\bar{\Xi}$\\
\>$\bar{k}N  \rightleftharpoons  k\Xi$
\> \>$k\bar{N}  \rightleftharpoons  \bar{k}\bar{\Xi}$\\
\>$\pi\Xi \rightleftharpoons k\Omega^- $
\> \>$\pi\bar{\Xi} \rightleftharpoons  \bar{k}\overline{\Omega^-}$\\
\>$k\bar{\Xi} \rightleftharpoons \pi\overline{\Omega^-}$
\> \>$\bar{k}\Xi \rightleftharpoons \pi\Omega^-$\\
\>$\bar{N}N$ annihilation\\
\>$\bar{Y}N$ annihilation\\
\end{tabbing}

where the hyperons are $Y$=$\Lambda$ or $\Sigma$. 
The relative probabilities 
for the different channels, e.g. in 
$(\pi$$N)$-
scattering, is used to determine
the outcome of the inelastic encounter. As the reactions introduced
above do not make up the full inelastic cross section, the remainder is
again treated as elastic encounters.

In LUCIAE\_2.0 a weight factor is given to each channel in order to
keep isospin conservation. In LUCIAE\_3.0 we do not need such a 
weight factor any more since the reverse reaction of each channel (except
$\bar{N}N$ and $\bar{Y}N$ annihilation) has been included. In this case
the inelastic cross sections used in the program should be 
looked upon as isospin-averaged cross sections and the cross sections 
of the reverse reactions are calculated by the detailed balance.

For $\bar{N}N$ and $\bar{Y}N$ annihilation its final states are simply
treated as the states of five particles through $\bar{N}N \rightarrow 
\rho \omega \rightarrow 5\pi $ and $\bar{Y}N \rightarrow K^*\omega 
\rightarrow K+4\pi$, respectively. The cross section of $\bar{Y}N$ 
annihilation is taken to be 1/5 of that of $\bar{N}N$annihilation.

\section{The Description of the Program}
\label{descript}
LUCIAE3.0 is a subroutine package for simulating collective gluon
emission in the Firecracker Model as well as the rescattering of the
produced particles from FRITIOF in a nuclear environment. The program
is written in FORTRAN 77. It should be used together with
FRITIOF7.02, \cite{H.p12}, JETSET7.4, \cite{jet}, PYTHIA 5.5, \cite{py} and
ARIADNE4.02, \cite{ari}. A user should read 
paper \cite{H.p12} first before using LUCIAE3.0 if he is not familiar
with using FRITIOF7.02. To run with LUCIAE3.0, FRITIOF7.02, JETSET7.4 and
ARIADNE4.02 have been somewhat modified and are called FRITIOF7.02R,
JETSET7.4R and ARIADNE4.02R, respectively.
One important thing that a user has to keep in mind is that the dimension 
size of the common block COMMON/LUJETS/ (KSZJ) and COMMON/FRPARAI (KSZ1) have 
been extended to KSZJ=40000 and KSZ1=30. Moreover, the space-time coordinates
of particles at the freeze-out time of the colliding system are now stored 
in array V(KSZJ,5). Those who have used FRITIOF7.02 before will find it is 
very easy to use FRITIOF7.02R which is linked to LUCIAE3.0 since the input and 
output structure of the two programs are completely the same. One simply 
starts with the subroutine FREVENT and ends with final state particles (after 
executing the Firecracker Model and rescattering) recorded in event record 
LUJETS. 

\subsection{User interface}
This section contains the most essential information a user should know about 
the program, namely the common block FRPARA1 and LUCIDAT2 for input
parameters and switch controls. The input parameters are provided with 
sensible default values. Several options are also included to make the program 
more accessible to user.

\begin{itemize}
\item{PARAMETER (KSZ1=30) \\
      COMMON/FRPARA1/KFR(KSZ1), VFR(KSZ1)}\\
\begin{description}
\item[Note:] KFR(1-14) and VFR(1-16) are used in FRITIOF7.02. We still list 
them here since some small changes have been 
made.
\item[KFR(1)] (D=1)
       Fragmentation
      \begin{description}
       \item[=0] Off.
       \item[=1] On.   
      \end{description}
\item[KFR(2)] (D=1) 
       Multiple gluon emission (dipole radiation) 
      \begin{description}
       \item[=0] Off.
       \item[=1] On.
      \end{description}
\item[KFR(3)] (D=0) Event selection for collisions with a nucleus 
      \begin{description}
       \item[=0] Generate minimum bias events (all interactions recorded).
       \item[=1] Generate only events with all projectile nucleons
                  participated.
       \item[=2] Generate only events with impact parameter between
                  $b_{min}$=VFR(1) and $b_{max}$=VFR(2).
       \item[=3] Apply both requirements in 1 and 2.
      \end{description}
\item[KFR(4)] (D=1)
       Fermi motion in nuclei 
      \begin{description}
       \item[=0] Neglected.
       \item[=1] Included.
      \end{description}
\item[KFR(5)] (D=0) 
       Nucleon-nucleon overlap function 
      \begin{description}
       \item[=0] Eikonal.
       \item[=1] Gaussian.
       \item[=2] Gray disc.
      \end{description}
\item[KFR(6)] (D=2)
       Target Nucleus deformation
      \begin{description}
       \item[=0] No deformation.
       \item[=1] Deformed target nucleus.
       \item[=2] Apply deformation only if the target atomic number $A\geq 108$.  
      \end{description}
\item[KFR(7)] (D=1) 
       Rutherford parton scattering processes
      \begin{description}
       \item[=0] Off.
       \item[=1] On. Here only the hardest RPS is used in FRITIOF.    
       \item[=2] On. The full multiple hard scattering scenario of PYTHIA
               is used.
      \end{description}
\item[KFR(8)] (D=1) 
      Hard gluons cause a corner (soft gluon kink) on the string
      \begin{description}
       \item[=0] No kink is formed. 
       \item[=1] Gluon kink is formed.
      \end{description}
\item[KFR(9)] (D=1)
        `Drowning' of Rutherford parton scattering 
      \begin{description}
       \item[=0] Off. Accept all RPS events.    
       \item[=-1] On. Throw away the drowned RPS event completely 
                  and replace it by a purely soft event.
       \item[=1] As in -1, but the 
            transverse momentum transfer of the soft collision is superimposed
            by the $q_T$ of the drowned RPS.  
      \end{description}
\item[KFR(10)] (D=1) 
        SRM parameters in RPS events: $\mu_1=\mu_0/r$, $\mu_2=\mu_0/(1-r)$
      \begin{description}
      \item[=0] $\mu$ remains the same as in a soft event: $\mu_1=\mu_2=\mu_0$.  
      \item[=1] $r$\,=\,VFR(16). 
      \item[=2] $r$ takes a uniform distribution in (0,1).
      \end{description}
\item[KFR(11)] (D=4)
      Write out of a message when the arguments in FREVENT is changed.  
      \begin{description}
      \item[=-1] Write it out every time the change occurs.    
      \item[=$n$ ($n\geq 0$)] The write out is limited to $n$ times.    
      \end{description}
\item[KFR(12)] (D=2) 
        Set up of the dipole cascade and string fragmentation parameters.
      \begin{description}
      \item[=0] No set up.  The default values are used. 
      \item[=1] Set to the values optimised by 
       OPAL collaboration, \cite{opal}:
       PARA(1)=0.20, PARA(3)=1.0, PARJ(21)=0.37, PARJ(41)=0.18, PARJ(42)=0.34.  
      \item[=2] Set to the values optimised by DELPHI collaboration:
       PARA(1)=0.22, PARA(3)=0.6, PARJ(21)=0.405, PARJ(41)=0.23, PARJ(42)=0.34.  
       \end{description}
\item[KFR(13)] (D=0) 
       Compresses the event record to save space in LUJETS.  This switch
       is particularly needed for heavy ion collisions at high energy
       where LUJETS must be compressed before it gets overfilled.  
      \begin{description}
      \item[=0] Do not compress LUJETS.  
      \item[=1-3] LUEDIT(KFR(13)) is called and LUJETS is compressed. 
        Specifically, for KFR(13)=1 fragmented jets and 
        decayed particles are removed, for KFR(13)=2 neutrinos and
        unknown particles are also removed, and for KFR(13)=3
        neutral particles are further excluded.
      \item[=4] A dummy subroutine FREDITD() is provided as an interface
        in which a user may write his own special purpose codes to edit
        and compress LUJETS.    
      \end{description}
\item[KFR(14)] (D=0) 
       Checked for charge and energy-momentum  conservation before rescattering
      \begin{description}
      \item[=1] The outcome of each event will be checked.
      \item[=0] No checked. 
       \end{description}       
\item[KFR(15)] (D=1) 
       FIRECRACKER model
      \begin{description}
      \item[=1] On.
      \item[=0] Off.
       \end{description}       
\item[KFR(16)] (D=1) 
       Firecracker gluon
      \begin{description}
      \item[=1] Accept firecracker gluons which are kinematically allowed.
      \item[=0] Reject all firecracker gluons. It is equivalent to KFR(15)=0.
       \end{description}     
\item[KFR(17)] (D=1) 
       The `drowning' of firecracker gluons by bremsstrahlung radiation.
      \begin{description}
      \item[=1] No drowning.
      \item[=0] Drowning.
       \end{description}     
  \item[KFR(18)] (D=1) 
         Size of firecracker cluster 
      \begin{description}
      \item[=1] The size of firecracker cluster = VFR(17)$<P_{t}>$,
         with $<P_{t}>$ being average transverse momentum of nucleons in the 
cluster.
      \item[=0] size of firecracker cluster =VFR(18).
       \end{description}   
\item[KFR(19)] (D=0) 
       Form of output, only used when KFR(21)=0
      \begin{description}
      \item[=1] Output in event record includes decayed strings.
      \item[=0] Decayed strings have been taken away in output.
       \end{description}       
\item[KFR(20)] (D=1) 
       The energy dependence of the cross section of an inelastic scattering
      \begin{description}
      \item[=1] Constant cross section.
      \item[=0] Energy-dependent cross section.
       \end{description}       
\item[KFR(21)] (D=1) 
       Rescattering
      \begin{description}
      \item[=1] On.
      \item[=0] Off.
       \end{description}     
\item[KFR(22)] (D=1) 
       Rescattering channel
      \begin{description}
      \item[=1] All channels are included.
      \item[=0] Only some channels of interest are included. The option is put
      here for future use.
       \end{description}     
\item[KFR(23)] (D=0) 
       Checked for charge and energy-momentum  conservation
        after rescattering
      \begin{description}
      \item[=1] The outcome of each event will be checked.
      \item[=0] No checked. 
       \end{description}  
\item[KFR(24)] (D=1)
         Treatment of spectators
	\begin{description}
	\item[=1] put  spectators  into a cluster. The cluster with the
	projectile spectators is placed in the (N-1)-th line of the event
	record with K(N-1,2)=10000+number of protons in the cluster and
	K(N-1,3)=number of neutrons in the cluster.The cluster with the
	target spectators is placed in the (N)-th line of the event
	record with K(N,2)=-10000-number of protons in the cluster and
	K(N,3)=number of neutrons in the cluster.
	\item[=0] leave spectators as individual nucleons.
	\end{description}
\item[KFR(25)] (D=1)
	         effect of effective string tension
	\begin{description}
	\item[=1] calculation of effective tension.
	\item[=0] no calculation of effective tension and JETSET default
	parameters are used.
	\end{description}
\item[KFR(26)] (D=0)
	   selection of a collision       
	\begin{description}
	\item[=1] select all the collision pairs.
	\item[=0] only select those collision pairs with cms energy larger enough
	to proceed one of the  inelastic reactions included.
	\end{description}

\vskip 12pt

\item[VFR(1)] (D=0.0 fm)
       Minimum impact parameter for options KFR(3)=2 or 3.
\item[VFR(2)] (D=0.2 fm)
       Maximum impact parameter for options KFR(3)=2 or 3.
\item[VFR(3)] (D=0.8 fm)
       The minimum allowable distance $R_{min}$ between nucleons in a nucleus.  
\item[VFR(4-5)] (D=0.2, 0.1)
       Dipole and quadrupole deformation coefficients for deformed target
       nucleus.
\item[VFR(6)] (D=0.01 GeV$^2/c^2$)
       The $<Q_T^2>$ for the Gaussian distribution of soft transverse 
       momentum transfer.
\item[VFR(7)] (D=0.20 GeV$^2/c^2$)
       The $<Q^2_{2T}>$ for the Gaussian distribution of primordial transverse
       momenta on the string ends.       
\item[VFR(8)] (D=0.75 GeV)
       Soft radiation coherence parameter $\mu_0$ for projectile hadron or
       nucleon. 
\item[VFR(9)] (D=0.75 GeV)
       Soft radiation coherence parameter $\mu_0$ for target hadron or nucleon. 
\item[VFR(10-11)] (D=0.0, 0.0 mb)
       Projectile-target nucleon total and elastic cross 
       sections, respectively.
       By default, they are taken from the parameterization 
       of Block and Cahn, \cite{block} (MSTP(31)=5 in PYTHIA).  
       The meson-nucleon cross sections are obtained
       simply by scaling down the Block-Cahn fit. The scale factor is
       $(2/3-a/\sqrt s)$, where $a=1.13$ GeV for pions and $a=3.27$ GeV for 
kaons
       are chosen to reproduce the low energy experimental data. For
       all the other baryons, it is treated as a pion if it is a meson
       and it is treated as a proton if it is a baryon.
       User may override the default by setting VFR(17-18) to positive values.
       However, the user assigned cross sections will only affect the
       N-N interaction probability in nucleus collisions. The
       probability for Rutherford parton scattering is not affected.  
\item[VFR(12)] (D=1.0 GeV/$c$)
       The $q_{Tmin}$ for Rutherford parton scattering.  
\item[VFR(13-15)] (D=1/6, 1/3, 1/2)
       The probabilities for assigning various spins and flavours to the diquark
       end of the string.  For example in a proton, VFR(13-15) are the
       probabilities of finding a $ud$ diquark of spin 1,
       a $uu$ diquark of spin 1, and a $ud$ diquark of spin 0, respectively. 
\item[VFR(16)] (D=0.5)
       The fraction $r$ in option KFR(10)=1. 
\item[VFR(17)] (D=1.0)
       The size of firecracker cluster with respect to the average transverse 
momentum of nucleons 
in the cluster.
       See also KFR(18).
\item[VFR(18)] (D=0.35GeV/c)
       The size of firecracker cluster when KFR(18)=0.
\item[VFR(19)] (D=6.5GeV)
       Minimal CMS energy for starting FIRECRACKER model.
\item[VFR(20)] (D=0.35mb)
      Cross section for  $\pi + N$ $\rightarrow{K+Y}$ when KFR(20)=1.
\item[VFR(21)] (D=3.0mb)
      Cross section for $\pi$ + $N$$\rightarrow$ $\Delta$+$\pi$ when KFR(20)=1.
\item[VFR(22)] (D=1.0mb)
       Cross section for $\pi$ + $N$$\rightarrow$ $\rho$+$N$ when KFR(20)=1.
\item[VFR(23)] (D=6.0mb)
       Cross section for $N$ + $N$$\rightarrow$ $\Delta$+$N$ when KFR(20)=1.
\item[VFR(24)] (D=3.5) the parameter $\alpha$ for calculating the effective string tension. 
\item[VFR(25)] (D=0.8) the parameter  $m_0$ for calculating the effective string tension.
\end{description}
\end{itemize}
\begin{itemize}
\item COMMON/LUCIDAT2/KFMAXT,PARAM(20),WEIGH(400)\\
\begin{description}
\item[KFMAXT] (D=32)
       The maximum number of particle species included in rescattering.
\item[PARAM(1)] (D=40.0mb)
       The total cross-section of reaction $NN$ .
\item[PARAM(2)] (D=25.0mb)
       The total cross-section of reaction $\pi$$N$  .   
\item[PARAM(3)] (D=35.0mb) 
       The total cross-section of reaction $KN$.    
\item[PARAM(4)] (D=10.0mb) 
       The total cross-section of  reaction $\pi\pi$.    
\item[PARAM(5)] (D=3.0mb) 
       The cross section of $\pi+\pi \rightarrow \overline{K}+K$.
\item[PARAM(6)] (D=0.85) 
       The ratio  of inelastic cross-section to total cross-section.
\item[PARAM(7)] (D=0.5fm) 
       The formation time of a hadron at its rest-frame.
\item[PARAM(8)] (D=0.01fm) 
       The minimal step of time elapsing.
\item[PARAM(9)] (D=0.1) 
       The accuracy of four-momentum conservation when sampling spectator 
nucleons.
\item[PARAM(10)] (D=2.0) 
       The size of effective rescattering region is PARAM(10)$R_{t}$ with 
$R_{t}$ being the
       radius of target.
\item[PARAM(11)] (D=0.16/fm$^3$) 
       The nucleon density of nucleus.
\item[PARAM(12)] (D=0.01 GeV$^2$$/c^2$) 
        $<P_{t}^2>$ for the Gaussian distribution of spectator nucleons.
\item[PARAM(13)] (D=28.0)
       $\bar{N}$ N annihilation cross section.
\item[PARAM(14)] (D=5.6)
       $\bar{\Lambda}$ N annihilation cross section. 
\item[WEIGH(1)-WEIGH(400)] not used.
\end{description}
\end{itemize} 
\subsection{Other common blocks}
The following common blocks are used for transmitting data internally. While 
they can be accessed to read out information, they should never be used to 
input data.
\vskip 12pt
\begin{itemize}
\item COMMON/FCRSOUT/ MCL(10),RCL(20),INEL(400),NST(2,150),ACL(2,150)\\
\begin{description}
\item[MCL(1): ]  
       The number of cluster from projectile.
\item[MCL(2): ]  
       The number of cluster from target.
\item[MCL(3): ]  
       The number of accepted firecracker gluons.
\item[MCL(4): ]  
       The number of the initial collision pairs of rescattering.
 \item [RCL(1): ]  
 JETSET parameter PARJ(1) averaged over all strings when KFR(25)=1.
  \item[RCL(2): ]  
  JETSET parameter PARJ(2) averaged over all strings when KFR(25)=1. 
  \item[RCL(3): ]  
   JETSET parameter PARJ(3) averaged over all strings  when KFR(25)=1.   
  \item[RCL(4): ]   
    JETSET parameter PARJ(21) averaged over all strings when KFR(25)=1.
  \item[RCL(5): ]  
  number of elastic rescatterings taken place in aggregate. 
  \item[RCL(6): ]  
   number of inelastic rescatterings taken place in aggregate.  
  \item[RCL(7): ]  
   number of inelastic rescattering that is treated as an elastic scattering
  in aggregate because the corresponding  inelastic channel is not 
  included in the program.
  \item[RCL(8): ]   
    number of total rescattering taken place in aggregate. 
\item[INEL(I): ]  
       The total number that I-th inelastic channel takes place in aggregate. 
Order can be found in the subroutine ``coinel'' in the source code.
                              
\item[NST(L,J): ] 
       The number of strings in J-th cluster. L=1,2 - index to label projectile 
(L=1) and target (L=2).
\item[ACL(L,J): ]  
       The invariant mass of J-th cluster. Index L has the same meaning as 
      in NST(L,J) . 
\end{description}
\end{itemize}
\section{Acknowledgments}
This work is partly supported by the national Natural Science Foundation of 
China and by the Postdoctoral Fund of China.
\newpage
\Large
\begin{center}
\bf {Appendix A}
\end{center}
\normalsize
The following is a sample main program.\\
\renewcommand{\thesection}{Appendix \Alph{section}}
\renewcommand{\theequation}{\mbox{A}\arabic{equation}}
\setcounter{equation}{0}
\begin{verbatim}
C..This program generates a few sample LUCIAE events, and then does
C..histogram for negatively charged particle multiplicity distribution
C..in O+Au collision at 200 GeV/nucleon lab energy.
C..This routine, loaded together with (FRITIOF_7.02R, ARIADNE_4.02R,
C..PYTHIA5.5 and JETSET7.4R) can be used to test the installation of programs.  
C---------------------------------------------------------------------

      PARAMETER (KSZJ=40000,KSZ1=30)
C...    **** Be sure to check that all the KSZJ's in MAIN, Fritiof,
C...         Jetset and Pythia are identically set      *****
      COMMON/FRPARA1/KFR(KSZ1),VFR(KSZ1)
      COMMON/FRINTN0/PLI0(2,4),AOP(KSZ1),IOP(KSZ1),NFR(KSZ1)
      COMMON/LUDAT1/MSTU(200),PARU(200),MSTJ(200),PARJ(200)
      COMMON/LUJETS/N,K(KSZJ,5),P(KSZJ,5),V(KSZJ,5)
      COMMON/LUDAT3/MDCY(500,3),MDME(2000,2),BRAT(2000),KFDP(2000,5)
      DIMENSION MP(0:300)

C...Open a file to take the write out of the program:
      MSTU(11) = 20
      OPEN(MSTU(11),FILE='test.out',STATUS='unknown')

C:::::::Multiplicity distribution for O+AU collision at 200 GeV ::::::::

C...Forbid the decays of Lambda and K_S0: 
      MDCY(LUCOMP(3122),1) = 0
      MDCY(LUCOMP(310),1) = 0

C...Book spaces for the histogram (or use a histogram package):
      DO 50 J=0,300
50    MP(J) = 0

C...Test 50 events (of course a lot more events are needed realistically):
      NEVENT=50
      NTRIG = 0
      DO 100 I=1, NEVENT

      CALL FREVENT('FIXT','O','AU',200.)

C...Output the event using JETSET routine LULIST:
      IF(I.LE.3) CALL LULIST(1)

C...Edit the event record, remove partons or decayed particles:
      CALL LUEDIT(1)

C...Assume a trigger requiring that the energy in the forward cone
C...(theta < 0.3 degree) must be less than 60% of the total beam energy.
C...Also find out the number of negatively charged particles:
      IQTRIG = 0
      EFWD = 0.
      N_=0
      DO 70 J=1, N
      THETA = PLU(J,14)
C...  (PLU is a JETSET function.  Please refer to the JETSET manual.)
      IF(THETA.LT.0.3) EFWD = EFWD+PLU(J,4)

C...Count the negative particles.  Spectator nucleons, which have codes
C...ABS(K(J,2))=10000+N_proton, must be excluded:
      IF(ABS(K(J,2)).LT.10000) THEN
        IF(PLU(J,6).LT.0.) N_ = N_+1
      ENDIF
70    CONTINUE

      EBEAM = 200.*IOP(3)
      IF(EFWD.LT.0.6*EBEAM) IQTRIG = 1

C...Do histogram:
      IF(IQTRIG.EQ.1) THEN
      NTRIG = NTRIG+1
      MP(N_) = MP(N_)+1
      ENDIF

100   CONTINUE

C...Output the histogram data:  
      WRITE(MSTU(11),500) NEVENT, NTRIG
      DO 200 J=0,300
200   WRITE(MSTU(11),*) J, FLOAT(MP(J))/FLOAT(NTRIG)
500   FORMAT(X,'Number of events:',I4,2x,'Triggered events:',I4)

C...Write out the values of the parameters and some statistics:
      CALL FRVALUE(0)
 		
      CLOSE (MSTU(11))
      END
\end{verbatim}

\end{document}